\documentclass[aip,amsmath,amssymb,reprint]{revtex4-1}

\usepackage{graphicx}
\usepackage{dcolumn}
\usepackage{bm}

\usepackage[utf8]{inputenc}
\usepackage[T1]{fontenc}
\usepackage{mathptmx}
\usepackage{color}

\draft 

\begin{document}


\title{Hydrogen Polarity of Interfacial Water Regulates Heterogeneous Ice Nucleation} 
\author{Mingzhe Shao}
\affiliation{Department of Packaging and Printing, Tianjin University of Science and Technology, Tianjin China}
\affiliation{School of Physical Sciences, University of Chinese Academy of Sciences, Beijing 100049 China}
\affiliation{Institute of Chemistry, Chinese Academy of Sciences, Beijing 100190, China}

\author{Chuanbiao Zhang}
\affiliation{School of Physical Sciences, University of Chinese Academy of Sciences, Beijing 100049 China}

\author{Conghai Qi}
\affiliation{Institute of Applied Physics, Chinese Academy of Sciences, Shanghai, China}

\author{Chunlei Wang}
\affiliation{Institute of Applied Physics, Chinese Academy of Sciences, Shanghai, China}

\author{Jianjun Wang}
\affiliation{School of Physical Sciences, University of Chinese Academy of Sciences, Beijing 100049 China}
\affiliation{Institute of Chemistry, Chinese Academy of Sciences, Beijing 100190, China} 

\author{Fangfu Ye}
\affiliation{School of Physical Sciences, University of Chinese Academy of Sciences, Beijing 100049 China}

\author{Xin Zhou$^{*}$}
\affiliation{School of Physical Sciences, University of Chinese Academy of Sciences, Beijing 100049 China}

\date{\today}

\begin{abstract}
Using all-atomic molecular dynamics(MD) simulations, we show that various substrates could induce interfacial water (IW) to form the same ice-like oxygen lattice but different hydrogen polarity order, and regulate the heterogeneous ice nucleation on the IW. We develop an efficient MD method to probe the shape, structure of ice nuclei and the corresponding supercooling temperatures. We find that the polarization of hydrogens in IW increases the surface tension between the ice nucleus and the IW, thus lifts the free energy barrier of heterogeneous ice nucleation. The results show that not only the oxygen lattice order but the hydrogen disorder of IW on substrates are required to effectively facilitate the freezing of atop water. 
\end{abstract}


\maketitle 

\section{Introduction} 
Water freezing on various material surfaces is of importance in wide-ranging fields~\cite{Hudait2018,Russo2014,Murray2012,Liou2000,Inada2012} such as cloud seeding in climate, frost heaving, and cell preserving. So far, the microscopic mechanisms of the key process of freezing on substrates, the heterogeneous ice nucleation, still remains elusive~\cite{Rausch2013,Moore2011,Kaufmann2017,Zhang2018}. It has been proposed that the lattice match of crystal surfaces of substrates to that of ice promotes water freezing, and has been regarded as the reason that silver iodide (AgI) greatly promotes ice nucleation~\cite{Vonnegut1947,Vonnegut1971}. However, the lattice matching mechanism is often questioned recently, since some materials with almost the same lattice to that of ice crystal, such as cuprous bromide (CuBr), barium fluoride (BaF$_{\mathrm{2}}$), do not facilitate ice nucleation~\cite{Swanson1953,Davies2014,Hu2007,Lupi2014,Reinhardt2014,Fraux2014,Zielke2014,Conrad2005,Chandler2013}. Besides, recent experiments and molecular dynamics (MD) simulations show many aspects of substrates, such as proton ordering~\cite{Sun2012}, charge~\cite{Gavish1992,Wilen1993}, sign of charge~\cite{Ehre2010}, electric field~\cite{Yan2012}, hydrophobicity~\cite{Cox2015,Cox2015b}, and the morphology of surfaces~\cite{Zhang2018,Fitzner2015}, can profoundly affect ice nucleation together, in cooperation or in competition. It is very hard to give a general picture to describe the mechanism of ice nucleation on these various surfaces. 

While the various kinds of aspects in different materials are hard to universally describe, all these different substrates induce atomic rearrangement of the first layers of interfacial water (IW), and regulate the freezing of water atop. It has found that the first layers of interfacial liquid on substrates rearrange the molecules and dominate supercooling~\cite{Schulli2010,Greer2010}, wetting~\cite{Wang2009,Kimmel2005,Zhu2013,Guo2015}, adsorption~\cite{Sun2012} and evaporating~\cite{Wan2015} of the atop liquids. It would be desirable to investigate the correlation between the atomic reconstruction of interfacial water (IW) and freezing of water on various kinds of substrates, which might provide an universal understanding about the effects of various kinds of substrates on ice nucleation. 

Recently, a number of MD simulations focusing on the oxygen rearrangement~\cite{Fraux2014,Zielke2014} and orientation~\cite{Abdelmonem2017,Sun2012}  of the IW have already revealed its significance. It was found that the oxygen atoms of the first layer of IW on $\beta-$AgI was almost perfect ice-like, and promoted the formation of ice-like lattice of next water layers thus freezing all of atop water. The orientation of IW was found to be crucial in heterogeneous ice nucleation as well, charge-induced $\alpha$-alumina surface suppresses ice nucleation upon it, irrespective of the sign of the surface charge. Moreover, the adsorption energy of polar monomer on the ice surface exhibits a strong correlation with the IW orientation. These recognitions emphasize the role of IW in heterogeneous ice nucleation, and inspires our interest in pursuing a more detailed perception of the specific effects of transitional and rotational order of IW. To our best knowledge, there is no report on hydrogen rearrangement influence on ice nucleation with oxygen lattice interference eliminated up to now. 

In this letter, based on all-atomic MD simulations, we found that the formation of ice-like oxygen lattice of IW alone is not sufficient to aid ice nucleation, the hydrogen polarization of IW induced by substrates also sensitively regulate freezing of atop water. This hydrogen dominating ice nucleation will generate fresh insight into the classical nucleation theory, and a broader perspective on the lattice matching mechanism as well.

 \section{Models} 
 \subsection{Simulation}  
We vary the ionic charge of neutral substrates with the lattice same as that of AgI to mimic the different charge of cation/anion in a series of different materials, such as AgI and CuBr, and investigate the reconstruction of IW on the substrates. 
We simulate, at most, $44800$ rigid TIP4P/Ice water molecules~\cite{Abascal2005,Vega2005} employing the GROMACS 4.6.2 software package. The model parameters of substrates are the same as that in reference~\cite{Zielke2014}, except the varied ionic charge, $q$. The NVT ensembles with Nose-Hoover thermostat and periodic boundary conditions (PBC) in only (x,y) directions are applied, while two resilient walls are induced in z-direction and far from water and substrates. Coulomb interaction is calculated based on the particle mesh Ewald method. The integrate time step is 2 fs.
For probing the freezing of water, we count the number of maximal ice cluster during $50~\mathrm{ns}$ regular MD simulation, following the method of Dellago and Doye~\cite{Lechner2008, Reinhardt2012,Russo2016a}, here a molecule is identified as ice or water based on its orientational order, and we group hydrogen-bonded ice molecules as clusters. 
 \subsection{The Proton Order} 
The regulated IW can have a different orientation even with the same oxygen lattice, there are 16 different unique H-bond ordering schemes possible in a unit ice I$_{h}$ cell\cite{Hirsch2004}. To identify these hydrogen ordering effects, an order parameter, $C_{OH}$, characterizing the arrangement of dangling OH bonds on the ice surface is defined in previous work as\cite{Sun2012}
\begin{eqnarray}
C_{\mathrm{OH}}=\frac{1}{N_{\mathrm{OH}}} \sum_{i=1}^{N_{\mathrm{OH}}} c_{i},
\end{eqnarray} 
This definition aims to identify the inhomogeneity of the ice surface. A larger $C_{OH}$ value means a more inhomogeneous ice surface, which generates an effective electric field, helping polar molecules to adsorb. However, it is somehow hard to link a practical material to this definition referred atom-scale inhomogeneity, In our case, the order parameter is defined to meet a more universal use as the hydrogen polarity of IW, 
\begin{eqnarray}
\xi= \frac{\sum_i \vec{p}_i \cdot {\hat n}}{\sum_i |\vec{p}_i \cdot {\hat n}|},  
\end{eqnarray} 
here $\vec{p}_i$ is the dipole moment of water molecule, $\hat n$ is the normal vector of the IW surface. $\xi=0$ refers to ordinary ice, the proton-disordered hexagonal ice phase ice I$_{h}$, while $\xi=1$ refers to ice XI, the proton-ordered form of ice I$_{h}$.
Another intuitive hydrogen polarity is defined to bring a direct recognition of the rotation of IW as below:
\begin{eqnarray}
\xi=\frac{\left|c_{1}+c_{2}-c_{4}-c_{5}\right|}{\sum_{k} c_{k}}.
\end{eqnarray} 

\begin{figure}[ht]
\centering{}\includegraphics[width=0.5\textwidth]{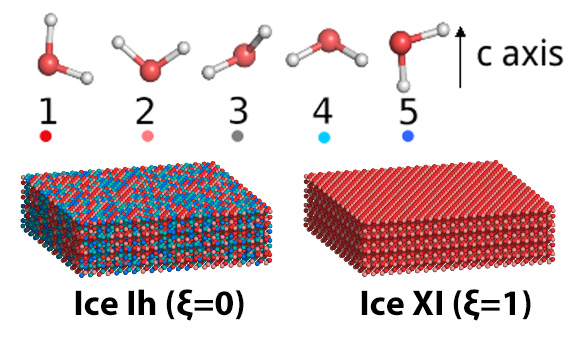} 
\caption{Five kinds of water molecules with different OH directions. A guide view for different ice structure with same lattice but different hydrogen polarity is also presented. 
\label{fig:figs1} }
\end{figure}

Here water molecules are divided into five kinds according to the direction of its OH bond, and the numbers (or concentrations) of the five kinds of water are denoted as  $c_k, k=1,…,5,$ respectively, see the Figure~\ref{fig:figs1}. For example, as the ordered form of Ice I$_{h}$, Ice XI $(\xi=1)$ only have type 1 and 2 water molecules. There is no significant statistical difference between these two definitions in our work.
This polarization of IW, which regulate ice nucleation upon it, can be very sensitive to external stimuli such as interface, charge and other force fields. To elucidate this process, we build the polarized IW in two different ways.

In AgI-like ($q=0.6e$ for AgI~\cite{Zielke2014})substrates, we verify the formation of the IW layer with almost perfect ice lattice, and the freezing of water on the surfaces within $10$ nanoseconds at $260~\mathrm{K}$. As increasing $q$ while keeping the substrate electrically neutral, the ice-like lattice of IW is more and more distorted and water becomes more difficult to freeze. To elucidate whether the ice-like oxygen lattice of IW is sufficient to facilitate the freezing of atop water, we apply soft harmonic springs to constrain the oxygen atoms of the first IW layer in their equilibrium positions in the AgI case to keep the ice-like lattice of the IW. The elastic constant of springs $k=50~ \mathrm{kcal/mol/}$\AA$^{2}$. For all applied $q$ from $0.3~e$ to $1.6~e$, the constraint can remains the ice-like lattice of the IW. In this case, the ice structure remains the same while hydrogen polarity increased from 0 to 0.3.

However, the polarity of IW upon AgI-like substrates is affected by atom charge, so could the nucleation behavior upon the IW. To exclude this atom charge effect, the IW is polarized by applying an electric field along the c-axis of ice I$_{h}$ lattice with a range from -0.8 to 1.2 eV/m at $260~\mathrm{K}$. For the same reason, a constraint on oxygen atoms is applied. The IW rotate and exhibiting hydrogen polarity from 0 to 0.7. It has to be aware of polarized IW and its corresponding ice structure is stable in bulk phase, or even possess lower internal energy comparable to the common Ice I$_{h}$. In our simulations, all polarized IW(from 0 to 0.7) keep its lattice and polarity unchanged when put into a periodic boundary condition as a bulk phase. Therefore the oxygen lattice is no longer account for the distinct heterogeneous nucleation upon polarized and non-polarized IW. This time, the hydrogen polarity effect could be the only reason count for the nucleation upon IW. We are looking forward to this study of heterogeneous nucleations would bring novel insights into classical nucleation theory.

 \subsection{Probing Critical Nucleus}  
It is difficult (if not impossible) to directly simulate water freezing on IW with larger $\xi$ due to the requirement of too long MD time. Sanz~\emph{et al.}~\cite{Sanz2013} developed an efficient indirect MD method to study the homogeneous ice nucleation of bulk water. They detected the corresponding supercooled temperature of a preset spheric ice nucleus instead of directly detecting the critical nucleus at a temperature. The main difficulty to expand the method in freezing on substrate surfaces is that the shape and structure of critical ice nucleus, such as the contact angles and crystalline surface of nucleus on substrates, are unknown. 

Here, we present a subtle simulation scheme to gradually adjust the shape and size of preset ice nucleus on substrates for approaching to a critical one, then get the corresponding supercooled temperature:    
(1) we generate a sphere-cap (or other shapes, such as spheric) hexagonal ice (I$_{h}$) nucleus and locate it on the surface then immerse them in supercooled water as the initial conformation. We also guess a few neighboring supercooled temperatures around a centra temperature; (2) we simulate from the initial conformation a segment of time, \emph{e.g.}, $10~ \mathrm{ns}$, at each set temperatures; (3) we choose one from these trajectories where the shape of ice nucleus was most obviously adjusted (but not completely melting out or growing up). The final conformation of chosen trajectory is set as new initial conformation, and the corresponding temperature is as the new centra temperature. We reset a few new simulation temperatures around the centra temperature by supposing the ice nucleus will grow or melt (with or without shape adjustment) at these temperatures, respectively. 
(4) We repeat the step (2) and (3) a few times, until the ice nucleus less correlates with the preset one at beginning, and its shape do not change obviously any more. Thus we get a critical ice nucleus on substrate, and we can get two neighboring temperatures where the ice nucleus grows and shrinks, respectively. The medium value approximately gives the corresponding temperature of the critical nucleus. 

\section{The Phase Diagram of Frozen Temperature and Hydrogen Polarity} 
 
Hydrogen polarity of IW, which measures the out-of-plane asymmetry of hydroxyl directions~\cite{Hirsch2004,Sun2012,Nada1997}, sensitively affects the formation of ice nucleus, even while the oxygens of IW form almost perfect ice-like lattice. On $\beta-$AgI-like surfaces, we adjust the charges of cation and anion (mimic different material with similar lattice).  For example, the surface with $q = 1.4~ e$ corresponds to BaF$_2$, (the other $F$ locates in the deeper position from the surface, is found to less affect the IW) while $q= 0.6~e$ for AgI~\cite{Fraux2014,Zielke2014}.

\begin{figure}[ht]
\centering{}\includegraphics[width=0.5\textwidth]{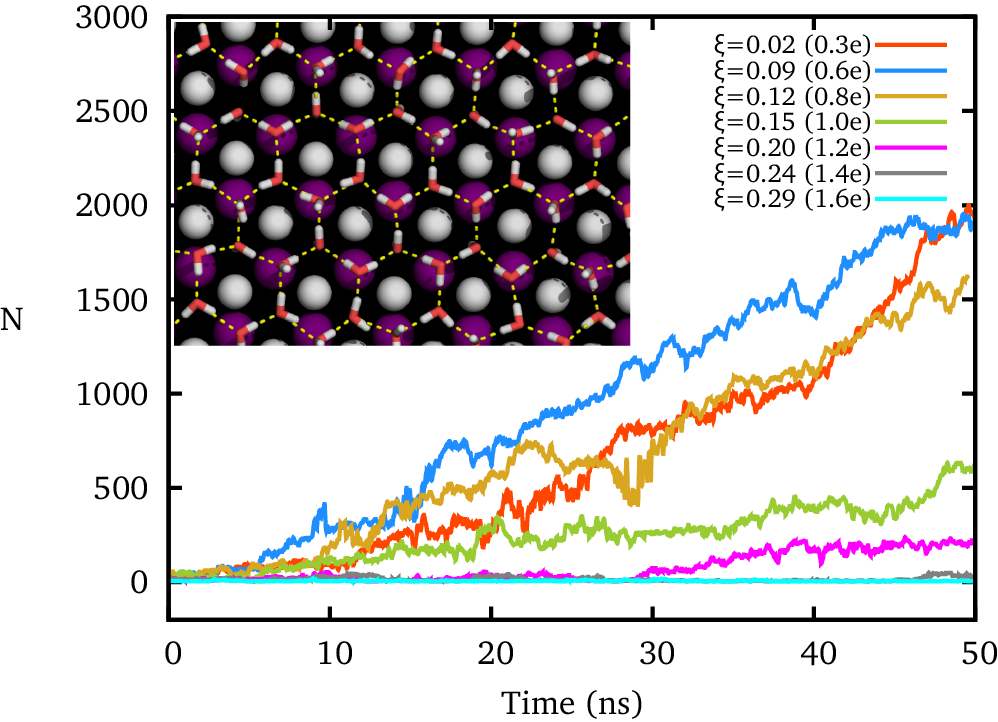} 
\caption{The growing of ice cluster upon the AgI-like substrates with different $q$ is shown. $T=260~\mathrm{K}$.
Inset: the ice-like first layer of interfacial water under constraint. Ag: white sphere; I: purple sphere. 
\label{fig:fig1} }
\end{figure}

\begin{figure}[ht]
\centering{}\includegraphics[width=0.5\textwidth]{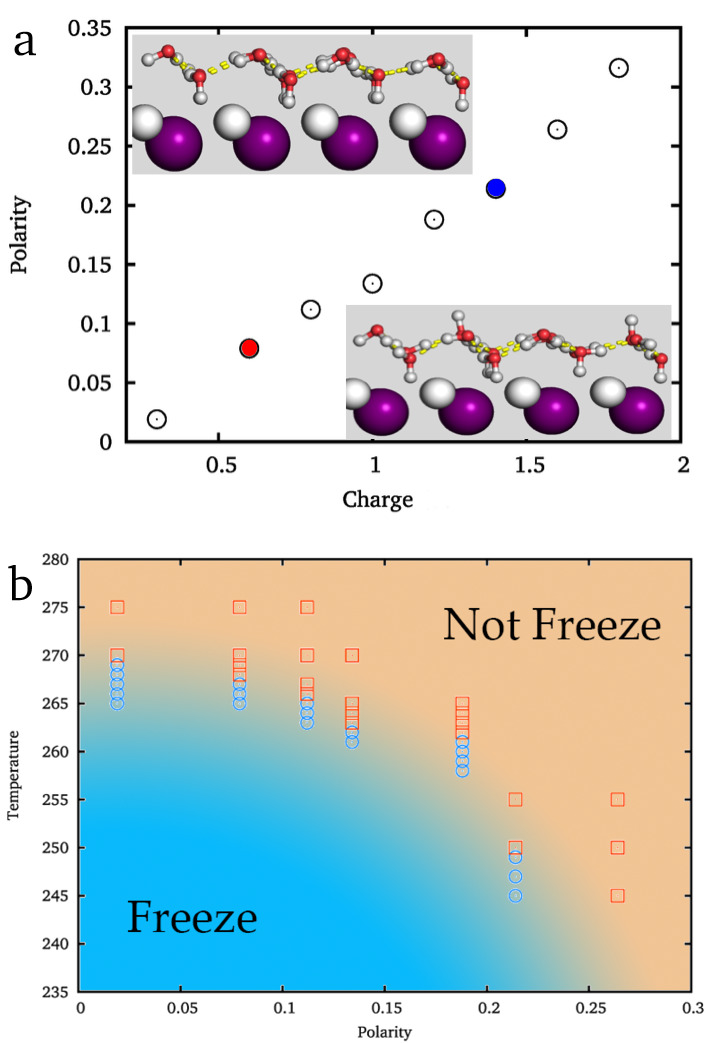} 
\caption{ (a) The hydrogen polarity of interfacial water varies as the ionic charge of surface. A local view revealing distinct difference between proton ordered and disordered ice on substrates, are shown while $q =0.6~ e$ (bottom-right) and $q= 1.4~e $ (top-left), . (b) The approximate  phase diagram of frozen temperature and hydrogen polarity of interfacial water. Each point is obtained by a $50~ \mathrm{ns}$ standard MD simulations from initial liquid.
 }
\label{fig:fig2}
\end{figure}

The freezing process upon these substrates are shown in Figure~\ref{fig:fig1}. 
For small $\xi$, an ice-like IW layer can quickly form with the ice I$_{\mathrm{h}}$-like lattice within $10$ nanoseconds at $260~ \mathrm{K}$ and facilitate the ice growth upon it, similar as the results on AgI surfaces~\cite{Fraux2014,Zielke2014}. 
However, as increasing $\xi$ by adjusting the ionic charge, the growing of ice cluster becomes slow. The ice-like lattice of IW is distorted on both the mimic BaF$_2$ with $q=1.4~ e$ and on the realistic BaF$_2$, water on top can not freeze within the $50~\mathrm{ns}$ at all, although the IW still has the same ice-like oxygen lattice due to constraints on oxygen atoms. This result is responsible for the less capability of BaF$_2$ on facilitating ice nucleation.

The main difference of IW in the different $q$ cases is its different polarity of hydrogen directions. As shown in the inset of Figure~\ref{fig:fig2}a, while $q=0.6~e$, the hydroxyl groups of IW almost equally point to the two sides of IW, but if $q=1.4~e$, more hydroxyls point toward the substrate and less hydroxyls point to the reverse direction, the water side. Therefore, the ice-like IW is hydrogen ordered~\cite{Hirsch2004,Sun2012,Nada1997} in large $q$ cases, but hydrogen disordered in small $q$ cases. 
When calculating the hydrogen polarity of IW, we find that $\xi$ monotonously increases from $0.02$ to about $0.29$ while $q$ increases from $0.3~e$ to $1.6~e$. 
By simulating $50~\mathrm{ns}$ each from initial liquid water at various temperatures, we get an approximate phase diagram of water freezing, see Figure~\ref{fig:fig2}b. 
The freezing temperature of water abruptly decreases as increasing $q$ of substrate thus $\xi$ of IW. While $q=0.3~e$ thus $\xi \approx 0$, water freezes at $269~\mathrm{K}$, approaching to the melting temperature of the applied TIP4P/Ice water model, $272~\mathrm{K}$. 
As increasing $q$ to $0.6~e$ (in AgI case), the hydrogen polarity of IW $\xi \approx 0.08$, water is found to freeze around  $265~\mathrm{K}$, and while $\xi \approx 0.21$ where $q = 1.4~e$ the frozen temperature decreases to about $250~\mathrm{K}$. 
 
\section{Critical Ice Nucleus upon Polarized IW} 

For detecting whether the rearrangement of IW or the ionic charge of the substrate itself controls  water freezing, we prepare the ice-lattice-like but polarized IW in the absence of (AgI-like) solid substrates, and check the water freezing on a few (usually $4\sim 6$) ice-like polarized IW layers. The polarized IW has the same ice lattice but gradually varied hydrogen polarity $\xi$ from $0$ (the ice I$_{h}$) to $1$ (the ice XI)~\cite{Salzmann2006,Salzmann2009,Raza2011,Fan2010,Geiger2014}.

\begin{figure}[ht]
\centering{}\includegraphics[width=0.5\textwidth]{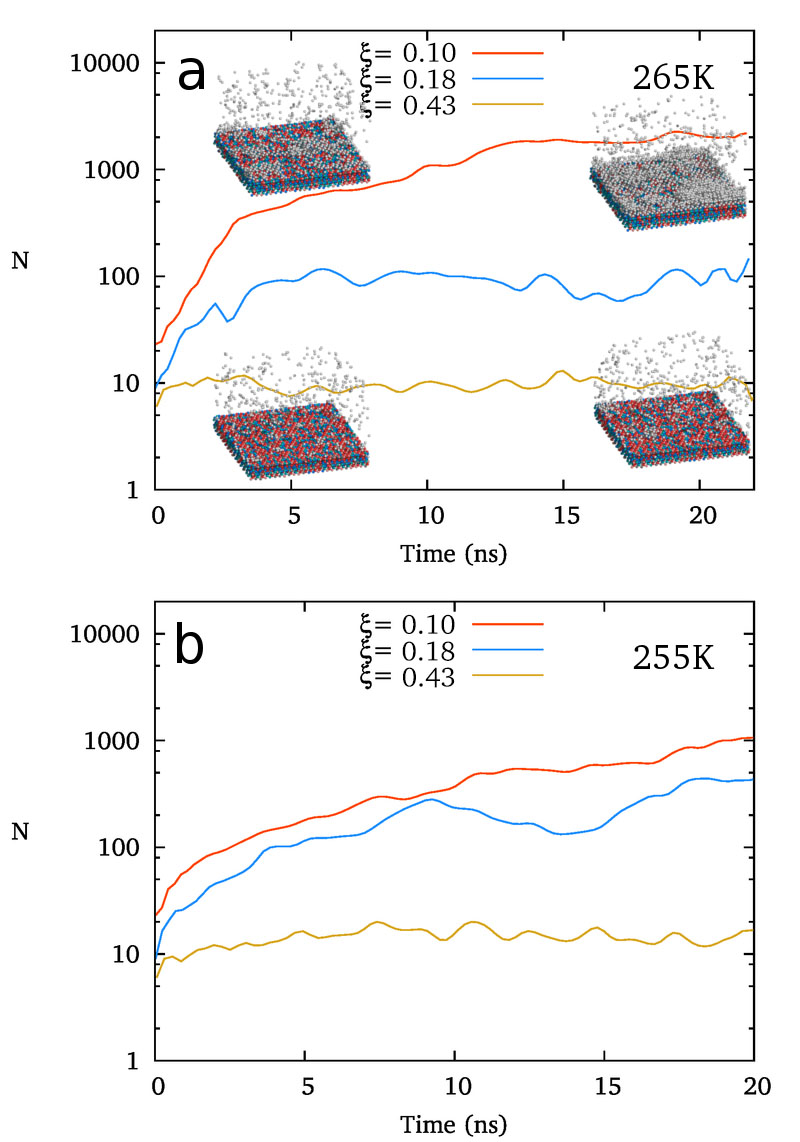} 
\caption{ The heterogeneous nucleation upon ice-like interfacial water at (a) 265K and (b) 255K. }
\label{fig:figs4}
\end{figure}

As shown in Figure~\ref{fig:figs4}, the freezing of supercooled water is also found to become more difficult as increasing the hydrogen polarity of the ice-like polarized IW in absence of AgI-like solid substrates.  
At $265~\mathrm{K}$, water freezes on the ice-like polarized IW layers with $\xi =0.10$, but does not form an ice cluster at $\xi=0.18$ during $20~\mathrm{ns}$ regular MD simulations. We also carried out simulations at $255~\mathrm{K}$, water freezes at $\xi=0.10$ and $\xi=0.18$, but not at $\xi=0.43$ within  $20~\mathrm{ns}$. This result indicates a significant heterogeneous nucleation barrier difference between IW of $\xi=0.10$ and $\xi=0.18$, which also matches the phase diagram of frozen temperature and hydrogen polarity of interfacial water we extract from AgI-like substrates.

\begin{figure}[ht]
\centering{}\includegraphics[width=0.5\textwidth]{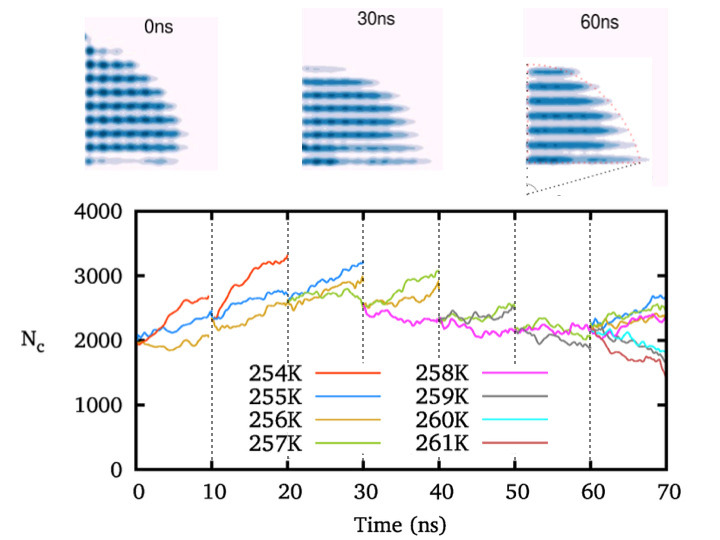} 
\caption{ The time evolution of the size and shape of ice nucleus during the simulations of adjusting critical ice nucleus  on the IW with $\xi=0.16$. The outlines of half ice nucleus at $0$, $30$ and $60~\mathrm{ns}$ are shown.  
\label{fig:fig3} }
\end{figure}

We apply the critical nucleus probing method on the ice-like polarized IW of various $\xi$ to get the critical ice nuclei and the corresponding temperatures $T$. Figure~\ref{fig:fig3} illustrates whole the simulation scheme on the ice-like IW with $\xi=0.16$. The ice nucleus is sufficiently adjusted to change its shape and its size (initial $2000$ to final $2350$ molecules) after $6 \times 10~\mathrm{ns}$ simulations. Then we simulate the final $10~\mathrm{ns}$ trajectories at a few temperatures and find the ice nucleus shrinks at $259~\mathrm{K}$ but growing at $258~\mathrm{K}$, thus we get the middle temperature $T=258.5~\mathrm{K}$ where the ice nucleus is in critical. 

\begin{figure}[ht]
\centering{}\includegraphics[width=0.5\textwidth]{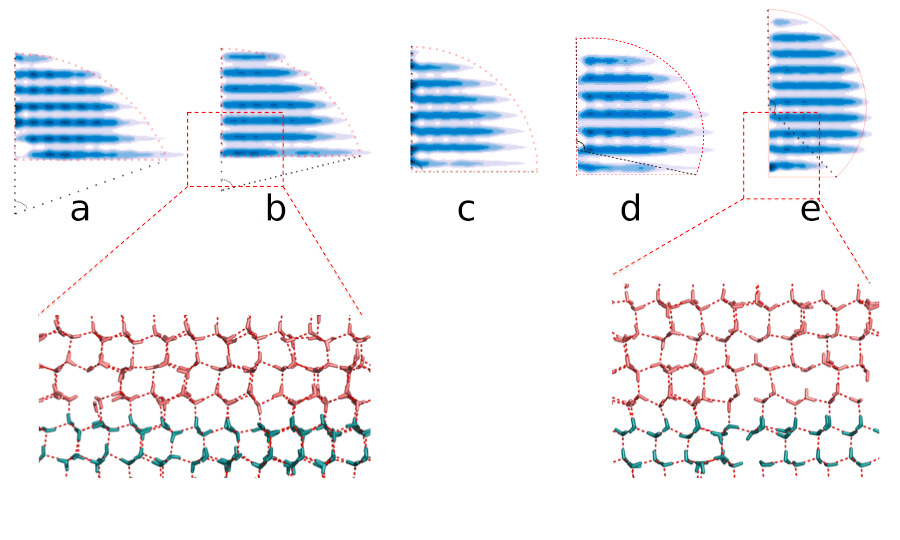} 
\caption{ Top: the obtained critical ice nuclei on the polarized IW surface with $\xi = 0.14, 0.16, 0.27, 0.39$ and $0.53$ in (a), (b), (c), (d)  and (e) respectively. All nuclei are sphere-cap, here only show the right half of them due to the symmetry. Bottom:  the hydrogen-bond connection between atoms in critical nucleus (red) and that in polarized IW (blue) with $\xi =0.16$ and $0.53$, respectively. 
\label{fig:fig4} }
\end{figure}

By extracting the outlines of the final ice nuclei from the average density of ice nuclei $\rho = 0.5 \rho_{I}$, we find that all the critical ice nuclei are approximately sphere-cap, as the expectation of the classical nucleation theory (CNT), except small deviation in the first layers in small and large $\xi$ cases, see Figure~\ref{fig:fig4}. Here $\rho_{I}$ is the density of bulk ice, about $0.906~ {\textrm g/cm^3}$ in this model. 

The critical temperatures of heterogeneous nuclei(about 2000 molecules) upon polarized IW of different levels is estimated as shown in Table~\ref{table:tab1}. What can be clearly seen in this table is the steady decline of critical temperature caused by the IW polarization.

\begin{table}
\caption{The obtained temperatures where the ice nuclei are at critical. After a few times of repeating, the shapes of clusters are effectively adjusted, their critical temperatures are then obtained.}

\centering{}%
\begin{tabular}{cccccccc}
\hline
{ repeat times} & {1} & {2}  & {3}  & {4} & {5} & {6} & {7}\tabularnewline
\hline
{ $\xi=0.14$} & { 263$\pm$2} & {263$\pm$1}  & {263$\pm$1}   & {262 $\pm$1} & {262 $\pm$1} & {262 $\pm$1} & {262 $\pm$1} \tabularnewline
{ $\xi=0.16$} & {$\sim$255} & {>256}  & {$\sim$257}   & {257 $\pm$1} & {258 $\pm$1} & {258 $\pm$1} & {258 $\pm$1} \tabularnewline
{ $\xi=0.27$} & {$\sim$255} & {>256}  & {$\sim$256}   & {257 $\pm$1} & {258 $\pm$1} & {257 $\pm$1} & {257 $\pm$1} \tabularnewline
{ $\xi=0.39$} & {$\sim$255} & {$\sim$255}  & {$\sim$255}   & {255 $\pm$1} & {255 $\pm$1} & {255 $\pm$1} & {255 $\pm$1} \tabularnewline
{ $\xi=0.53$} & {$\sim$255} & {$\sim$255}  & {$\sim$254}   & {$\sim$254} & {253 $\pm$1} & {252 $\pm$1} & {252 $\pm$1} \tabularnewline
\hline
\end{tabular}
\label{table:tab1}
\end{table}

From the simulations, we have the size $N_c$, radius $R$, the (apparent) contact angle $\theta$ of the sphere-cap critical nucleus, the corresponding (supercooled) temperature and the free energy barrier $\Delta G$ of nucleation, shown in Table~\ref{table:tab2}.

\begin{table}
\caption{The critical ice nuclei on the ice-like IW with different hydrogen polarity $\xi$. The unit of radius $R$ is the interlayer distance of ice, about $3.7$\AA, (the error of R is about $0.1$ in the unit); free energy $\Delta G$ in $\mathrm{kcal/mol}$.}

\centering{}%
\begin{tabular}{cccccccc}
\hline
{ $\xi$} & {$\theta$} & { $f\left(\theta\right)$}  & {R}  & {$T_c$(K)} & {$\Delta G$ }\tabularnewline
\hline
{ 0.14} & { 72} & {0.28}  & {10.0}   & {262 $\pm 1$} & {50}\tabularnewline
{ 0.16} & { 78} & {0.35} & { 8.8}   & {258 $\pm 1$}& {32} \tabularnewline
{ 0.27} & { 90} & {0.50 }  &{ 7.7}  & {257 $\pm 1$}  & {40} \tabularnewline
{ 0.39} & { 98} & {0.60 }  &{ 6.9}  & {255 $\pm 1$}  & {37} \tabularnewline
{ 0.53} & {134} & {0.94}  &{ 6.1}   & {252 $\pm 1$} & {42}\tabularnewline
\hline
\end{tabular}
\label{table:tab2}
\end{table}

We estimate the free energy barriers from the CNT, $\Delta G = f(\theta) \frac{4 \pi \gamma}{3 (\kappa \Delta T)^{2}}$. Here $f(\theta)=\frac{1}{4}(2+\cos\theta)(1-\cos\theta)^2$ is the shape factor of spherical cap. 

\begin{figure}[ht]
  \centering
\includegraphics[width=0.5\textwidth,angle=0]{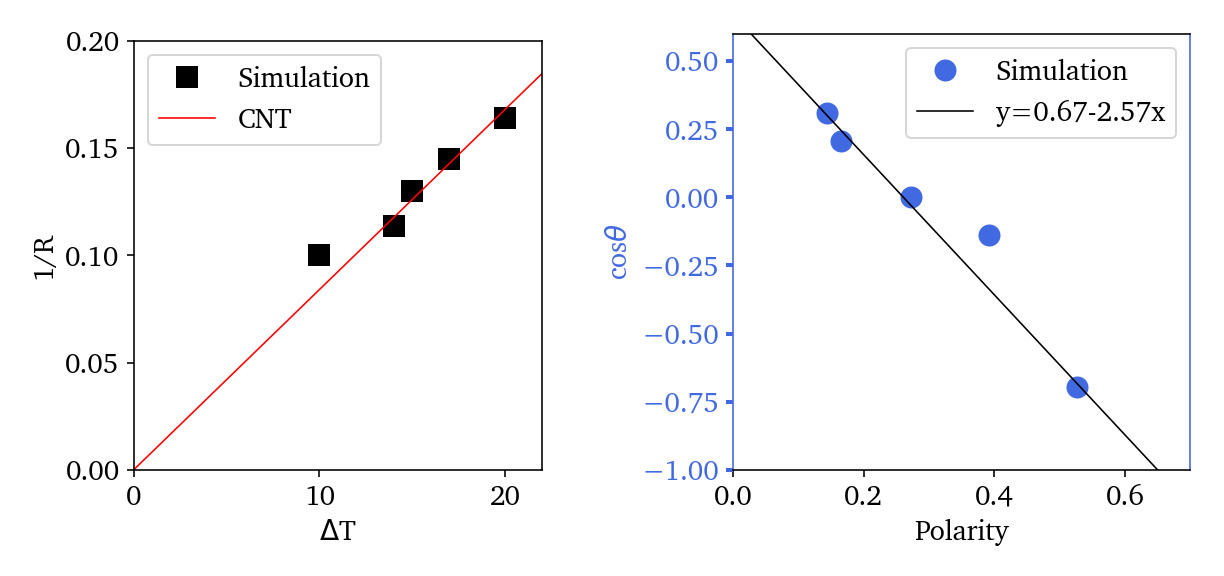}
\caption{ Left, the reverse radius of ice nuclei versus supercooling temperatures. The line is fitted based on the classic nucleation theory; right, the relation between contact angles of ice nuclei versus the hydrogen polarity of IW. 
\label{fig:fig5} 
}
\end{figure}

In Figure~\ref{fig:fig5}, we find that the reverse of radius $R$ of critical nucleus is proportional to the corresponding  supercooled temperature, $1/R \approx \kappa \Delta T$, with 
$\kappa \approx 0.03~ \mathrm{nm^{-1} K^{-1}}$. The result is in good agreement with the expectation of CNT, and $\kappa = \frac{\rho_{I} \alpha}{2 \gamma} \approx 0.02~ \mathrm{nm^{-1} K^{-1}}$. Here 
$\alpha = \frac{|\Delta \mu|}{\Delta T}$ about $0.0043~\mathrm {kcal/mol/K}$~\cite{Sanz2013}, the ice-water surface tension $\gamma \approx 26~\mathrm{mN/m}$, and $\Delta \mu$ is the chemical potential difference between ice and water. 
 
The cosine of contact angle is found to be linearly related to $\xi$ in the whole range $0<\xi<1$, as 
shown in Figure~\ref{fig:fig5}. 
From the Young's equation, we have 
\begin{eqnarray}
\frac{\Delta \gamma(\xi)}{\gamma} \approx 0.67 - 2.57 \xi.
\end{eqnarray}
Here $\Delta \gamma(\xi) = \gamma_{water,IW}(\xi) - \gamma_{ice, IW}(\xi)$, the difference of the water-IW and the ice-IW surface tensions. 

Considering the fact that the IW with $\xi = 0$ is similar to the normal hexagonal ice, 
we have $\gamma_{ice,IW}(\xi) = \gamma (\delta_1 + k_1 \xi + \cdots)$, while 
$\gamma_{water,IW}(\xi) =\gamma (1- \delta_2 + k_2 \xi +\cdots)$. Here both $\delta_1$ and $\delta_2$ are small positive values, and $k_1 > k_2 > 0$, since liquid water is more flexible than ice nucleus to rearrange its conformations on IW. Therefore, we have, $\delta_1 + \delta_2 \approx 0.33$, and $k_1 - k_2 = 2.57$. The higher order dependence of surface tensions on $\xi$ seems very small (or cancel each other) even while $\xi$ is approached to unity, where the completely polarized IW distorts the lattice of both itself and the growing ice nucleus to avoid dangling hydrogen bonds. 
  
\section{Conclusions} 
As summary, we show that the matching between the structure of interfacial water (IW) and the ice, involving both the ice-like oxygen lattice order and the hydrogen direction disorder, corresponds to the capability of substrates on the heterogeneous ice nucleation, only the lattice matching of substrates with ice may be not sufficient to aid ice nucleation. The result is helpful to finding and designing anti-/aid- freezing materials for application. 

\section*{Acknowledgement} 
The work is under the financial support of the NSFC Grant with No. 11574310, 11674345, 21733010. C.-L. Wang thanks the support of the Youth Innovation Promotion Association, CAS. M.-Z. Shao thanks the support of Beijing National Laboratory for Molecular Sciences (BNLMS).


\bibliography{horder}
\bibliographystyle{aipnum4-1} 

\end{document}